# Passive circular Brownian motion of asymmetric particles weakly bound to a planar surface

Guilherme H. Oliveira, A. Honorato, and Rene A. Nome[*]
*Institute of Chemistry, State University of Campinas (UNICAMP), Campinas, SP, Brazil, 13083-970*

**Abstract:** We use a simple model of particle shape to investigate how particle asymmetry affects particle-surface interaction, orientation, and stochastic dynamics over a planar surface. With this geometric model, we construct potential energy curves as a function of particle orientation relative to the surface, and identify the potential energy minimum for particles of various shapes ranging from symmetric (sphere) to asymmetric (oval-shaped). The calculated difference between particle centroid position and potential energy minimum location is used to define an offset, which is useful for comparison with experimental particle trajectories. For asymmetric particles the potential energy minimum location is decoupled from the center of the particle long-axis. Based on these observations, we construct a Brownian motion model of a rigid rotor with one end fixed to the planar surface. In the case of asymmetric particles, the resulting stochastic trajectories exhibit passive circular Brownian motion, thus providing one possible microscopic mechanism for the stochastic dynamics of fluorescence upconversion nanoparticles near a surface previously reported by us.

## 1. Introduction

Within the context of biophysical and molecular biology, fluorescence upconversion nanoparticles (UCNP) have the potential to enable characterization of microrheological properties in over a broad dynamic range and at high spatial resolution with optical microscopy and laser tweezers[1-3]. Thus, with the goal of exploring novel applications of UCNP in biophysical investigations, we have recently reported our experiments and simulations of stochastic dynamics of fluorescence upconversion nanoparticles (UCNPs) in the presence of thermal, harmonic, optical and non-conservative forces[4,5]. In references [4,5], the UCNP were characterized by high-resolution scanning electron microscopy, x-ray diffraction, dynamic light scattering, and zeta potential measurements. On average, the UCNP particles exhibited spheroidal shape with 600 nm size and +20 mV zeta potential. Single UCNP measurements were performed by fluorescence upconversion micro-spectroscopy and optical trapping. The mean-squared displacement (MSD) from single UCNP exhibited a time dependent diffusion coefficient, which we compared to Brownian dynamics simulations of a viscoelastic model of harmonically bound spheres. Moreover, experimental time-dependent two-dimensional trajectories of individual UCNP revealed correlated circular Brownian motion of individual nanoparticles. These trajectories were compared with stochastic trajectories calculated using the above viscoelastic model in the presence of a nonconservative rotational force field. Although rotational Brownian motion has been reported previously in various setting [6-14], our work described in [4,5] highlighted the complex interplay of UCNP shape, adhesion and thermal fluctuations that led to rich stochastic dynamics of these nanoparticles.

We sought to investigate whether nanoparticle orientation over the substrate could help us understand our experimental results reported in [4,5]. We consider oval-shaped asymmetric particles whose cross-section can be described by two axes (Figure 1)[15,16]. At equilibrium, the major axis symmetric objects such as an ellipse will be parallel to the surface, whereas the major axis in oval-shaped objects will be oriented at a finite small angle with respect to the surface in order to minimize the overall system free energy. For the same reason, the minor axis will be collinear with the vertical axis for an ellipse and tilted with respect to the vertical axis for oval-shaped objects. Furthermore, such orientation of major and minor axis with respect to horizontal and vertical planes will increase as the particle asymmetry increases.

---

[*] Corresponding Author: nome@iqm.unicamp.br

Based on this hypothesis, we have determined particle parameters that would allow us to explore the role of nanoparticle asymmetry in determining its orientation over a planar surface. We calculate potential energy with respect to the surface, and define the particle orientation to be that which minimizes potential energy. The minimum potential energy particle orientation is used to construct a rotational Brownian motor model. The resulting trajectories highlight circular motion in the case of rotation by fixed, rigid bodies. The calculated trajectories are consistent with the trajectories observed experimentally [4,5].

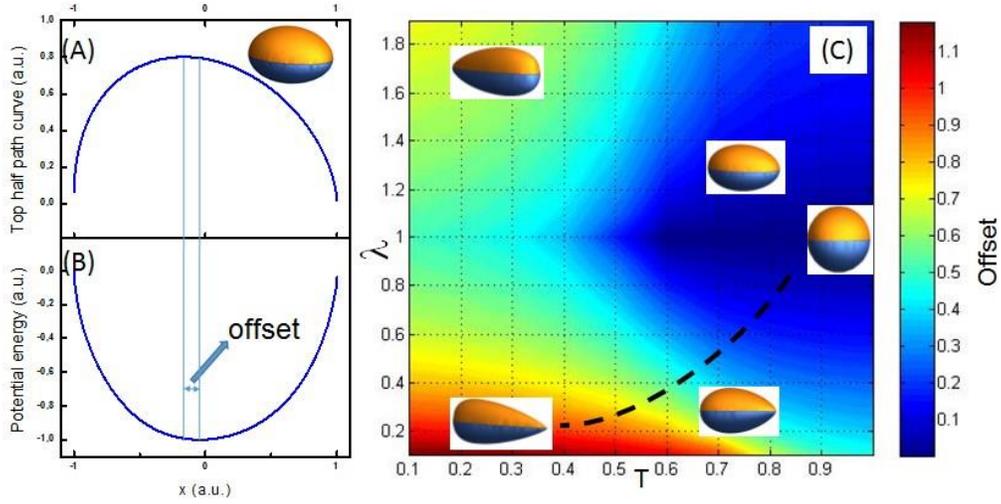

**Figure 1: Particle orientation relative to planar surface as a function of shape. (A) Top half path curve calculated using equation (1) with λ =1.39 and T = 0.79. Inset: 3D view of the particle with the same parameters. (B) Normalized potential energy as a function of orientation relative to the surface. "offset" indicates the difference between potential energy minimum and particle centroid. (C) Offset across the λ-T parameter space investigated herein. The inset shows particle shapes calculated using (λ,T) at the locations specified in the text, and the black dashed line indicates the range of parameter chosen for further analysis.**

## 2. Materials and Methods

To model the particle shape, we employ the equation of a path curve for the top half of an oval-shaped particle sized to a unit circle [15,16]:

$$y = T(1+x)^{\frac{1}{1+\lambda}}(1-x)^{\frac{\lambda}{1+\lambda}} \qquad (1)$$

Three-dimensional solid bodies were constructed by applying rotational matrix transformation around the particle major axis. We determine particle centroid by the density weighted arithmetic average assuming homogeneous particles with uniform density. Potential energies are calculated using the above equation since cross-sections suffice to determine particle orientation over the planar surface (Similar conclusions are reached when calculating the potential energy assuming a solid body with homogeneous density and uniform sampling – not shown). We treat the path curve as a discrete and homogeneous distribution with constant charge density such that the total potential energy at the planar surface at any given point is the sum of energies due to each discrete portion. The potential energy calculation is repeated for 2000 particle orientations relative to the planar surface. The closest distance between particle to the surface - consistent with the definition of Zeta potential - is kept constant for different particle orientations and shapes such that a direct comparison between systems can be made. The orientation space explored ranges from base (rounded end) to tip (pointed end). The equilibrium configuration is that which minimizes the potential energy of the system. We define the offset to be the difference between the equilibrium configuration corresponding to the potential energy minimum and the center of mass position (see Figure 1B for an example $\lambda$ = 1.39 and T = 0.79). Figure 1C is a plot of such offset as a function of the ($\lambda$,T) parameter range mentioned above.

We model two-dimensional Brownian motion of a rigid body by solving the Langevin equation of motion for a freely rotating particle using [17]:

$$\gamma \frac{d\theta}{dt} = \frac{1}{m} \delta F(t) \qquad (2)$$

where $\theta$ is the particle orientation, m is the particle mass, $\gamma$ is the friction coefficient, $\delta F(t)$ is a random force with zero mean and variance given by the corresponding fluctuation dissipation relation:

$$\langle \delta F(t) \delta F(t') \rangle = 2\gamma k_B T \delta(t-t') \qquad (3)$$

where $k_B$ is Boltzmann's constant.

## 3. Results and discussion

Figure 1A(inset) shows a 3D model of the asymmetric particle calculated using equation as described in Materials and Methods with T = 0.79 and $\lambda$ = 1.39. Figure 1A(main) shows the path curve for the particle top half calculated using T = 0.79 and $\lambda$ = 1.39. As shown in Figure 1A (main and inset), this choice of parameters illustrate particle asymmetry, whereby the center of mass coincides with the geometric center (the body is assumed to be homogeneous) yet the orthogonal minor axis is offset from long axis mean (set to zero in this work). Spherical, ellipsoidal and oval-shaped particles can be calculated by varying these two parameters while maintaining the major axis length fixed, as shown in the inset of Figure 1C. Thus, we can study the effect of particle asymmetry on its preferred orientation when deposited over a substrate.

Although particle shape calculation was performed using arbitrary units, experimentally, the UCNP particles described in [4,5] were 600 nm in size (hydrodynamic diameter). From the Stokes-Einstein-Debye equation for rotational Brownian motion in water at room temperature, elongated particles with 600 nm long axis length exhibit rotational diffusion times of nearly 200 ms and thus so-called "circling frequency" of 5 Hz [6]. Near planar surfaces and including electrostatic attraction, rotational dynamics ought to slow further. Thus, assuming the shaped objects shown in Figure 1 have characteristic length scales as the UCNP particles and that the measured Zeta potential (+20 mV) is on the order $k_BT$ at room temperature, we expect to observe circular Brownian motion as in references [4,5].

The potential energies were calculated by describing both particle and the surface as sets of point charges, as described in the Materials and Methods section, assuming the particle is positively charged (Zeta potential = +20 mV) and negatively charged planar surface (pKa of silanol groups found in optical microscope cover slip range from 4.5 to 7). By varying the relative orientation of the particle with respect to the surface, we calculate the potential energy curve, as shown in Figure 1B for $\lambda = 1.39$ and $T = 0.79$. The difference between potential energy minimum position and the geometric center is termed offset and reflects the stable nanoparticle orientation that maximizes its attraction towards the planar surface. Although particle orientation relative to the surface is a simple and intuitive concept, we prefer to use the term "offset" as it corresponds to the observable quantity in quantitative optical microscopy centroid-based particle tracking algorithms.

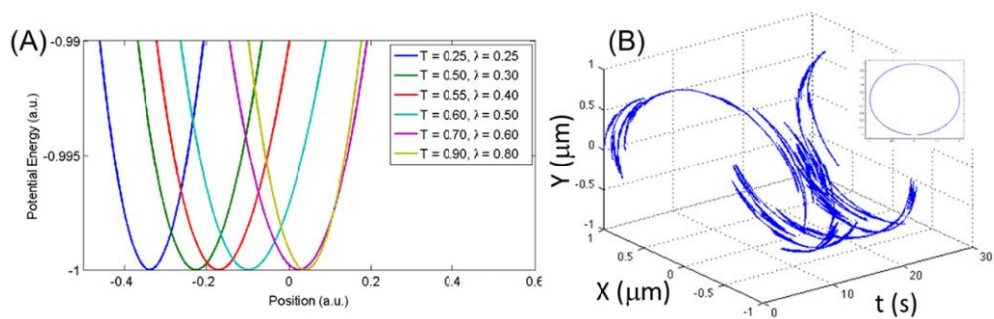

**Figure 2. Energetics and Brownian motion of asymmetric particles adhered to planar surface. (A) Potential energy as a function of orientation for different particle shapes along the dashed line shown in Figure 1C. (B) Circular Brownian motion of twodimensional rigid rotor fixed to a surface at the potential energy minimum. Simulation parameters: particle size = 1 $\mu$m; T = 300 K; viscosity $\eta$ = 0.001 N s / m$^2$**

Symmetric particles such as sphere ($\lambda = 1$ and $T = 1$) and ellipse (for example, $\lambda = 1$ and $T = 0.9$) have zero offset; thus, geometric center and potential energy minimum position are the same as expected. On the other hand, the offset between potential energy minimum and the geometric center increases as the particle asymmetry increases. For example, in the case of $\lambda = 1.39$ and $T = 0.79$ (as in Figure 1A), the offset is x = 0.01 unit length; in this case, the minimum energy configuration corresponds to the particle oriented with respect to the planar surface at an angle of approximately 5° degrees, which is typical of average oval-shaped objects (e.g. bird eggs [15,16]).

As shown in Figure 2A, depending on the specific parameters $\lambda$ and $T$, the offset can be zero or as large as half the particle long axis length, which is consistent with our experimental observations [4,5]. The parameter set chosen for potential energy calculation shown in Figure 2A follows the dashed line in Figure 1C. Thus, Figure 2A shows potential energies particle shape ranging from symmetric to asymmetric. In particular, for ellipsoidal-like shapes (T = 0.9 / $\lambda = 0.8$ and T = 0.7 / $\lambda = 0.6$), the particles are oriented such that the major axis is parallel to the planar surface, the potential energy minimum is located near the long axis center (x = 0). On the other hand, as particle asymmetry increases (T = 0.6 / $\lambda = 0.5$; T = 0.55 / $\lambda = 0.4$; T =

0.5 / $\lambda$ = 0.3 and T = 0.25 / $\lambda$ = 0.25 in Figure 2A), the particle major axis orientation is tilted relative to the planar surface. As a result, the potential energy minimum is located farther from the long axis center as the particle asymmetry increases.

Assuming the particle is weakly bound to the surface (due to an interplay of thermal and electrostatic forces) at the potential energy minimum, the distance between the potential energy minimum shown in Figure 2A and the long axis center can be used to define a projection of the long axis on the planar surface. Thus, we model stochastic dynamics of the asymmetric particles described here as Brownian motion of a rigid rotor in the rotating frame, as discussed in the Materials and Methods section. This model assumes the particle is bound more strongly to the surface at the potential energy minimum. Given the asymmetric particle shape, particle-surface interactions are weaker away from the potential energy minimum. The combination of particle asymmetry, stronger adhesion at the minimum and thermal forces is thus described as a rigid Brownian rotor with one end fixed to the surface. Figure 2B shows the calculated two-dimensional trajectories for an offset of 1.0, which corresponds to the parameter range $\lambda$ = 0.25 and T = 0.1 – 0.7 (see Figure 1C). The inset shows the calculated trajectory projected onto the X-Y plane. We note that the simulation results shown in Figure 2B do not include rotational force field terms as in our previous work. Thus, we believe the model presented here provides a reasonable microscopic basis for the circular Brownian motion observed in references [4,5]. On the other hand, in the case of more symmetric particles such that the offset is smaller than the spatial resolution of the optical microscope (i.e., less than ~ 0.5 µm), then the particle trajectories might be modelled by the viscoelastic model described previously.

## 4. Conclusions

We use a simple model of particle shape to investigate how particle asymmetry affects particle-surface interaction, orientation, and stochastic dynamics over a planar surface. With this geometric model, we construct potential energy curves as a function of particle orientation relative to the surface, and identify the potential energy minimum for particles of various shapes ranging from symmetric (sphere) to asymmetric (oval-shaped). The calculated difference between particle centroid position and potential energy minimum location is used to define an offset, which is useful for comparison with experimental particle trajectories. For asymmetric particles the potential energy minimum location is decoupled from the center of the particle long-axis. Based on these observations, we construct a Brownian motion model of a rigid rotor with one end fixed to the planar surface. In the case of asymmetric particles, the resulting stochastic trajectories exhibit passive circular Brownian motion, thus providing one possible microscopic mechanism for the stochastic dynamics of fluorescence upconversion nanoparticles near a surface previously reported by us.

## 5. Acknowledgments

Financial support from CNPq, CAPES, FAPESP and SAE-UNICAMP is gratefully acknowledged.

## 6. References

1. T.G. Mason and D.A. Weitz, "Optical measurements of frequency-dependent linear viscoelastic moduli of complex fluids," Phys. Rev. Lett. **74**, 1250-1253 (1995).
2. Z. Cheng and T.G. Mason, "Rotational diffusion microrheology," Phys. Rev. Lett. **90**, 018304 (2003).
3. S. Parkin, G. Knoner, W. Singer, T.A. Nieminen, N.R. Heckenberg, and H. Rubinsztein-Dunlop, "Optical microrheology using rotating laser-trapped particles," Methods Cell. Biol. **82**, 525-561 (2007).

4. R.A. Nome, B. Barja, C. Sorbello, M. Jobbágy, and V. Sanches, "Stochastic dynamics of fluorescence upconversion nanoparticles", Optics in the Life Sciences Congress, OSA Technical Digest, paper OtW2E.2, doi.org/10.1364/OTA.2017.OtW2E.2 (2017).
5. R.A. Nome, C. Sorbello, M. Jobbágy, B.C. Barja, V. Sanches, J.S. Cruz, and V.F. Aguiar, "Rich stochastic dynamics of co-doped Er:Yb fluorescence upconversion nanoparticles in the presence of thermal, nonconservative, harmonic and optical forces", Meth. App. Fluor. **5** (1), 014005 (2017).
6. S. Jahanshahi, H. Lowen, and B. ten Hagen, "Brownian motion of a circle swimmer in a harmonic trap," PhysRevE, **95**, 022606 (2017).
7. M. De Corato, F. Greco, G. D'Avino, and P.L. Maffettone, "Hydrodynamics and Brownian motions of a spheroid near a rigid wall,", J.Chem.Phys. **142**, 194901 (2015).
8. F. Kummel, B. Ten Hagen, R. Wittkowski, I. Buttinoni, R. Eichhorn, G. Volpe, H. Lowen, and C. Bechinger, "Circular motion of asymmetric self-propelling particles," Phys. Rev. Lett. **110**, 198302-1-5 (2013).
9. B. Gutiérrez-Medina, A.J. Guerra, A.I.P. Maldonado, Y.C. Rubio, J.V.G. Meza., "Circular random motion in diatom gliding under isotropic conditions," Phys. Biol. **11**, 066006 (2014).
10. O. Brzobohaty, A.V. Arzola, M. Siler, L.Chvátal, P. Jakl, S. Simpson, P. Zemanek, "Complex rotational dynamics of multiple spheroidal particles in a circularly polarized, dual beam trap", Optics Express, **23** (6), 7273 (2015).
11. H.E. Ribeiro and F.Q. Potiguar, "Active matter in lateral parabolic confinement: from subdiffusion to superdiffusion," Physica A **462**, 1294 (2016).
12. O.S. Duarte and A.O. Caldeira, "Effective coupling between two Brownian particles", Phys. Rev. Lett. **97**, 250601 (2006).
13. H.H. Wensink, V. Kantsler, R.E. Goldstein, J. Dunkel, "Controlling active self-assembly through broken particle-shape symmetry," Phys. Rev. E **89**, 010302(R) (2014).
14. P. Figliozzi, N. Sule, Z. Yan, Y. Bao, S. Burov, S.K. Gray, S.A. Rice, S. Vaikuntanathan, and N.F. Scherer, "Driven Optical Matter: Electrodynamically coupled nanoparticles in an optical ring vortex," Phys. Rev. E **95**, 022604 (2017).
15. M.C. Stoddard, E.H. Yong, D. Akkaynak, C. Sheard, J.A. Tobias, and L. Mahadevan, "Avian egg shape: form, function, and evolution," Science **356**, 1249-1254 (2017).
16. D. Baker, "A geometric method for determining shape of bird eggs," Auk **119**, 1179-1186 (2002).
17. G. Volpe, S. Gigan, G. Volpe, "Simulation of the active Brownian motion of a microswimmer", Am. J. Phys. **82** (7), 659 (2014).